\documentclass[%
 reprint,
 amsmath,amssymb,
 aps,
]{revtex4-2}

\usepackage{graphicx}
\usepackage{dcolumn}
\usepackage{bm}

\usepackage{xcolor}

\newcommand{\sout}[1]{}
\newcommand{\rev}[1]{\textcolor{black}{#1}}
\newcommand{\revtwo}[1]{\textcolor{black}{#1}}

\begin{document}

\preprint{APS/123-QED}

\title{Extreme Spatial Dispersion in Nonlocally-Resonant Elastic Metamaterials}

\author{Aleksi Bossart}
\email{aleksi.bossart@epfl.ch}
\author{Romain Fleury}%
\email{romain.fleury@epfl.ch}
\affiliation{Laboratory of Wave Engineering, Ecole Polytechnique Fédérale de Lausanne, 1015 Lausanne, Switzerland}%

\date{\today}

\begin{abstract}
To date, the vast majority of architected materials have leveraged two physical principles to control wave behavior, namely Bragg interference and local resonances. Here, we describe a third path: structures that accommodate a finite number of delocalized zero-energy modes, leading to anomalous dispersion cones that nucleate from extreme spatial dispersion at 0 Hz. We explain how to design such zero-energy modes in the context of elasticity and show that many of the landmark wave properties of metamaterials can also be induced at an extremely subwavelength scale by the associated anomalous cones, without suffering from the same bandwidth limitations. We then validate our theory through a combination of simulations and experiments. Finally, we present an inverse design method to produce anomalous cones at desired locations in \rev{k-space}.
\end{abstract}

\maketitle



In traditional elastic metamaterials, local resonances modify the spectrum by introducing a flat band which hybridizes with the background dispersion, thereby opening a polariton band-gap \cite{zhou_elastic_2012}. When two such gaps overlap (Fig.\ref{fig1}(ab)), one having negative effective density \cite{chan_extending_2006} and the other negative bulk modulus \cite{huang_theoretical_2011}, a \textit{doubly-negative} range arises, in which waves propagate with opposite phase and group velocities \cite{schuster_introduction_1924, lamb_group_1904, huang_anomalous_2012}. This property is associated to spatial dispersion \cite{forcella_causality_2017,belov_strong_2003} and provides a mechanical realization of Veselago's vision of negative refraction \cite{Veselago_1968, pendry_negative_2000}. While metamaterials were initially confined to electromagnetism \cite{pendry_extremely_1996, pendry_magnetism_1999}, they have since also flourished in acoustics \cite{liu_locally_2000,fang_ultrasonic_2006,yang_membrane-type_2008}, thermal physics \cite{fan_shaped_2008} and mechanics \cite{christensen_kadic_kraft_wegener_2015,bertoldi_flexible_2017,yu_flexural_2006,kane_topological_2014, meeussen_geared_2016}, merging with auxetic materials \cite{lakes_foam_1987,milton_which_1995}. \rev{Long-range interactions have recently been shown to extend the capabilities of metamaterials, either in the form of radiative coupling \cite{zhou_broadband_2022} or with explicit beyond-next-neighbor connectors \cite{chen_roton-like_2021,iglesias_martinez_experimental_2021,chaplain_reconfigurable_2022}. Such couplings induce local minima in the dispersion branches, called rotons \cite{landau_helium, feynman_energy_1956}.}

\begin{figure}[!ht]
\includegraphics[width=\columnwidth]{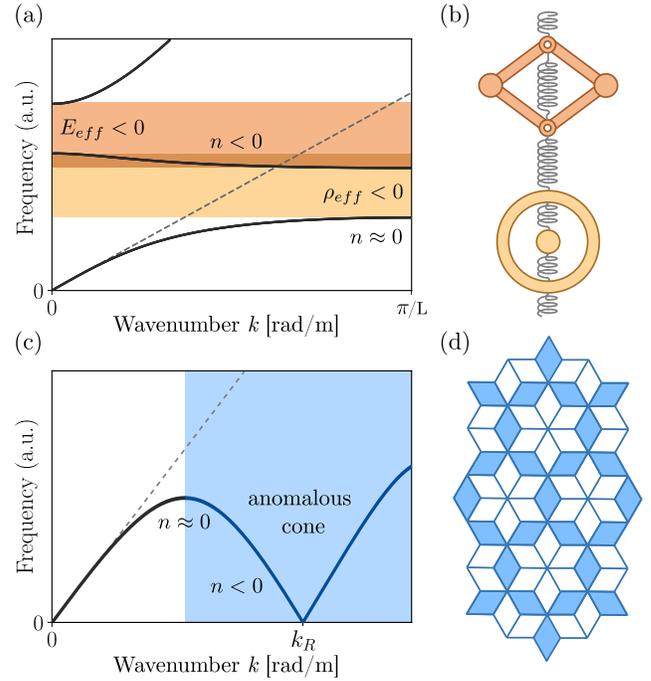}
\caption{\label{fig1} (a) Dispersion of a locally-resonant metamaterial
, with the ranges of negative bulk modulus and density highlighted in orange and \rev{yellow}, respectively. The narrow domain in which they overlap (\rev{brown} region) hosts a band with negative group velocity. The lower band corresponds to a polariton, whose subwavelength non-local dispersion is also controlled by the local resonances. (b) 1D spring-mass chain giving rise to the dispersion in (a), with the mechanisms responsible for effective negative compressibility and density colored accordingly. (c) On the other hand, the dispersion of a nonlocally-resonant metamaterial features a domain exhibiting an anomalous cone (\rev{blue} area). This cone is due to the presence of a delocalized mode with finite wavenumber $k_R$ at 0 Hz. (d) A geometry that hosts such nonlocal elastic resonances down to 0Hz.}
\end{figure}

\rev{Here, we consider a design path that relies neither on local resonances nor on long-range coupling, while inducing hallmark properties of wave metamaterials over large frequency bands. Our inspiration comes from the particularly intriguing \textit{interlaced wire media}} \cite{chen_metamaterials_2018,sakhno_longitudinal_2021}. These architected materials consist in collections of space-spanning, fully-connected meshes of metallic wires that interlace with each other without direct contact, thereby allowing for differences in static electric potential between their large disconnected components. The connectivity of the interlacing pattern can induce anomalous dispersion cones at symmetric points of the Brillouin zone \cite{chen_metamaterials_2018}; as a result, interlaced wire media exhibit unusual electromagnetic wave properties down to the static regime, such as broadband negative refraction.

In this Letter, \rev{we identify the abstract ingredients} underlying the exotic physical properties of such wave media, which we propose to call \textit{nonlocally-resonant metamaterials}. \rev{Indeed, they essentially rely on delocalized eigenmodes at zero frequency, or \textit{nonlocal resonances}, to parallel the traditional terminology.} This new paradigm allows us to expand their scope to the elastic realm, providing a theoretical inverse-design procedure and an experimental validation. 

In elasticity, \rev{a delocalized zero-frequency mode} consists in a locally-rigid deformation (or \textit{mechanism}) that spans the whole medium. \rev{More precisely, we require this resonance to} have a well-defined wavevector $\mathbf{k_R}$, isolated in \rev{k-space}. Indeed, if the immediate neighborhood of $\mathbf{k_R}$ also hosted zero-energy modes, we could construct zero-frequency wavepackets\rev{,} which would make the resonance local. In contrast, an isolated $\mathbf{k_R}$ guarantees sample-spanning spatial extension in the form of a Bloch wave. Crucially, the continuity of the spectrum ensures that an anomalous cone nucleates from $\mathbf{k_R}$, as in Fig.\ref{fig1}(c). \rev{Here, ``anomalous" refers to dispersion cones not associated to the translation symmetries which underlie the standard longitudinal and transverse elastic waves.} A trailblazing geometry that fits these stringent requirements- is the counter-rotating squares structure, introduced to model displacive phase transitions in minerals \cite{giddy_determination_1993, dove_theory_1997} through the rigid-unit mode description. Similar mechanisms were also described in a topological study of the deformed square lattice \cite{rocklin_mechanical_2016}. Here, we provide a general paradigm to understand and inversely design extremely nonlocal elastic resonances. We explore theoretically and experimentally the intriguing wave physics of nonlocally-resonant elastic metamaterials.


Let us begin by considering the metamaterial of Fig.\ref{fig1}(d), and picture it as a collection of oscillating masses \rev{(blue diamonds)} connected by springs (dark bars). In zero-frequency oscillations, only rigid motions are possible; not a single spring can be stretched. Rigid translations of the entire structure \rev{fulfill} this criterion; such zero-frequency modes correspond to transverse and longitudinal elastic waves in the limit of infinite wavelength, which confines them to the $\Gamma$ point. In a structure with high connectivity, these are the only zero-energy modes. Removing a sufficient number of rigid springs makes it possible for localized rigid motion to occur, typically leading to Guest-Hutchinson modes \cite{guest_determinacy_2003,lubensky_phonons_2015}, which have 1D extension both in real space and \rev{k-space \cite{Note1}}. Such modes cannot create anomalous dispersion cones, which are pinned to isolated zero-frequency points in \rev{k-space}. Instead, we \rev{seek} a geometry hosting a finite number of mechanisms\rev{, and therefore a finite number of anomalous cones. In addition, our ultimate goal of inversely designing such cones necessitates a geometry complex enough to be tailorable. We therefore need to go beyond the classical counter-rotating squares geometry and turn to oligomodal metamaterials \cite{bossart_oligomodal_2021}}. Indeed, oligomodal geometries are defined through their property of hosting a finite number of zero-energy modes that does not scale with increasing system size. \rev{We must add a second requirement, namely that such modes be of the Bloch-wave form, with well-defined $\mathbf{k_R}$}. This is the case of the particular metamaterial of Fig.\ref{fig1}(d), as we now demonstrate.

\begin{figure}[]
\includegraphics[width=\columnwidth]{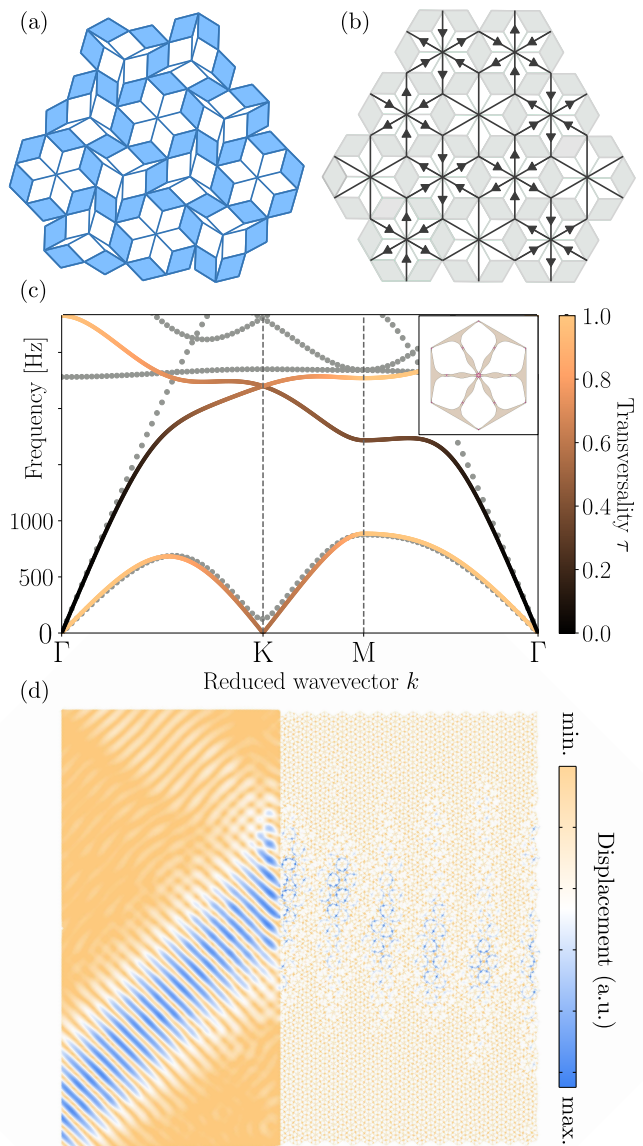}
\caption{\label{fig2}(a) A zero-energy deformation mode of the structure of Fig.\ref{fig1}(d). (b) Corresponding graph, \rev{overlaid on the undeformed geometry.} (c) Dispersion relation along symmetry lines of the Brillouin zone, with \rev{solid } lines corresponding to a spring-mass model and \rev{grey} dots to a finite-element simulation with the geometry depicted in the inset. \rev{The spring-mass spectrum is colored according to the transversality of corresponding eigenmodes.} (d) Finite-element simulation of a monochromatic gaussian beam negatively refracted at the interface between an isotropic material \rev{(left)} and this nonlocally-resonant metamaterial \rev{(right)}.}
\end{figure}

\rev{To do so, we use a convenient} description of mechanisms in terms of directed graphs \cite{bossart_oligomodal_2021}. We will first use \rev{these graphs} as a visual tool, and later unleash their full potential to inversely design nonlocally-resonant metamaterials. \rev{Step-by-step examples are detailed in \cite{Note1}.} In this method, one starts by drawing a graph with a vertex in every empty polygon enclosed by rigid springs (also known as a \textit{linkage}, in mechanical engineering). A second type of vertex is drawn on every point mass; we will refer to such vertices as \textit{hinges}. We then draw edges connecting every linkage vertex to the hinges surrounding it. To describe non-trivial rigid motions, arrows are drawn on each edge to depict relative rotation of the two neighboring rigid springs, with the number of arrows being proportional to the angular deformation. From a mechanical structure, such as the one shown in Fig.\ref{fig2}(a), we obtain a directed graph encoding angular deformations, such as the one of Fig.\ref{fig2}(b). Importantly, such deformations only entail local rotations and occur without stretching, consistently with the zero-frequency character of the modes we seek.

Kinematics impose some constraints on the allowed arrow configurations. First, all vertices are subject to arrow conservation, meaning that the number of incoming arrows must exactly balance the number of outgoing arrows; this reflects the fact that the sum of angles is conserved both for linkages and hinges. Second, the vertices associated to linkages are subject to additional constraints encoding their mechanical degrees of freedom; for instance, vertices surrounded by three rigid springs are statically determined and can be erased along with their edges. Vertices of degree four possess one degree of freedom, which means that fixing the arrow count on one edge determines the other three. In general, vertices of degree $n$ have $n-3$ degrees of freedom. The precise arrow-drawing rule derives from a linearization of the trigonometric relations between the angles of the linkage \cite{bossart_oligomodal_2021} and can be derived from symmetry considerations in simple cases. \rev{In general, this leads to vertices with unequal arrow weights, possibly non-integer.} Therefore, the problem of finding non-trivial modes at zero frequency becomes combinatorial \cite{coulais_combinatorial_2016, meeussen_topological_2020}; one needs to find all linearly independent arrow configurations that respect the aforementioned rules. 

The directed graph of Fig.\ref{fig2}(b) \rev{was obtained by applying the combinatorial rules to the structure of Fig.\ref{fig1}(d);} there, the unit-cell graph has been simplified to an equivalent vertex of degree six. This simplified vertex has one degree of freedom, which is drawn as alternating incoming and outgoing arrows of equal intensity. The combinatorial game goes like this: we draw an arrow on an arbitrary edge, say in the top-left cell of Fig.\ref{fig2}(b). This fixes five other edges through the kinematic vertex rule, but the remainder of the graph is not yet fully determined. We therefore need to choose the direction of another arrow in a neighboring cell, keeping in mind that incoming and outgoing arrows must compensate each other at any given vertex. In particular, this fixes the third arrow at any vertex shared by two neighboring unit cells and therefore determines the arrow distribution on that unit cell as well. We encourage the reader to try it out and see that the arrows then propagate over the whole graph coherently, yielding Fig.\ref{fig2}(b). Under the corresponding deformation mode, the structure of Fig.\ref{fig1}(d) is deformed into the one of Fig.\ref{fig2}(a). Since fixing two arrows was sufficient to reach that result, we see that the lattice has two mechanical degrees of freedom that act system-wide; they correspond to two elastic nonlocal resonances. By continuity, we therefore expect the spectrum to host two anomalous cones, located at the wavevectors associated to the two zero-energy modes we identified. Consider their spatial periodicity, \rev{easily observed} on the graph of Fig.\ref{fig2}(b); the deformation pattern repeats only after three cells in the horizontal direction. Since we can select the second mode as a mirror image of the first, our graph theory predicts anomalous cones that nucleate at zero frequency from the $K$ and $K'$ points of the Brillouin zone.


In order to verify this, we let the springs breathe a little, allowing for oscillations at non-zero frequencies. Collecting the displacements of all point masses in a vector $\mathbf{u}$, we can describe such oscillations through the differential equation $\partial_t^2\mathbf{u}=\mathcal{D}\mathbf{u}$, where the dynamical matrix $\mathcal{D}$ computes the effect of displacements on springs and restitutes the force the latter exert on point masses \cite{lubensky_phonons_2015}. Connecting masses on opposite boundaries of the unit cell with phase-shifted springs to implement Floquet-Bloch boundary conditions, we then compute a phononic band structure for a spring-mass model with the geometry of Fig.\ref{fig1}(d), with \rev{blue} diamonds replaced by equivalent triangulated spring frames \cite{Note1}. The result is depicted by the solid lines of Fig.\ref{fig2}(c); as expected, an additional cone emerges from the $K$ point. 

Let us discuss the impact of these anomalous cones on waves. First, they induce domains of negative group velocity that extend from 0Hz over a large bandwidth. Wavepackets prepared in that region of \rev{k-space} would be negatively refracted at an interface with an isotropic elastic medium. As for wavelengths closer to $\Gamma$, they would be positively refracted and split in longitudinal and transverse components, as usual. \rev{This phenomenon may be leveraged to filter wavelengths. If only the anomalous cone is required, one can instead pin the unit cell, thereby removing the cones near $\Gamma$ by destroying the underlying translation symmetry \cite{Note1}.} At higher frequencies, a second feature stands out: the anomalous cone hybridizes with another one. For instance, in Fig.\ref{fig2}(c), it connects with the cone of transverse elastic waves; momenta near the inflection point feel a very low index of refraction. Interestingly, this cone hybridization also creates a partial band-gap in which only longitudinal waves are allowed. We can go further with band gaps. Indeed, pre-twisting the metamaterial, for instance by selecting Fig.\ref{fig2}(a) as the rest position, can induce full band gaps \footnote{See Supplemental Material at [URL] for
\rev{(I) Introduction to the directed graph method; (II) Impact of pre-twisting on band structure; (III) Bandwidth of negative refraction; (IV) Prototyping details; (V) Details of inverse design method.}}. Remarkably, we identified gaps with relative bandwidth up to $\frac{\Delta\omega}{\omega}=56\%$, which is made possible by their low-frequency character. Such graph-preserving pre-twists also impact the speed of sound associated to anomalous waves, potentially tuning it all the way down to a flat band \cite{Note1}. These examples illustrate that as a general rule, the vertex model only fixes the location of the cones, while the precise shape and hybridization of the bands can vary a lot for geometries sharing the same zero-frequency graph.

\begin{figure}[]
\includegraphics[width=\columnwidth]{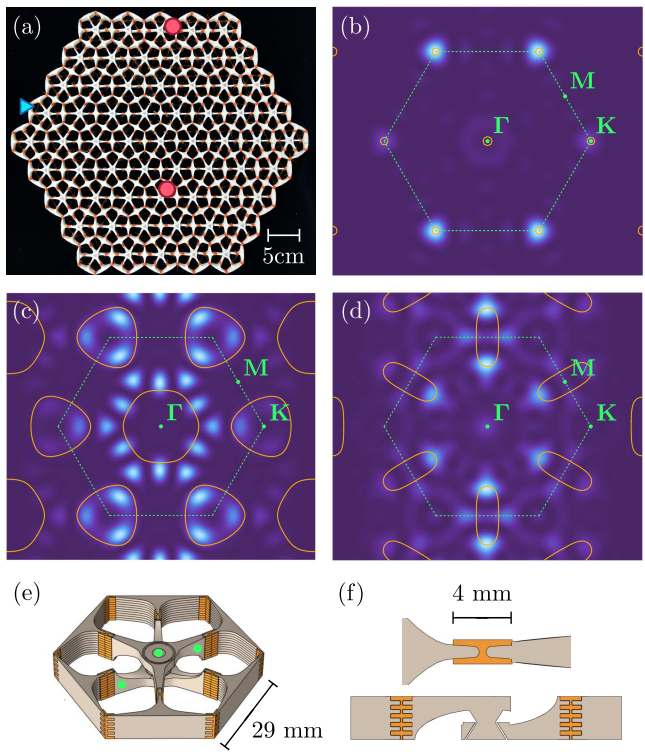}
\caption{\label{fig3} Experimental validation. (a) 3D-printed sample, with fixed points denoted by red circles and excitation points by blue triangles. (b-d) Spatial Fourier transform of $|\Omega|^2$ at $106Hz$, $529Hz$ and $806Hz$, respectively. Solid yellow lines represent the isofrequency contours (first band) predicted by the spring-mass model. Green dashed lines mark the limits of the first Brillouin zone, and symmetry points are labeled. \rev{(e) Geometry of the 3D-printed unit-cell, with PLA regions in brown and TPU regions in orange. Measurement points are indicated in green. (f) Closeup of the hinge geometry and cross-section of the unit-cell.}}
\end{figure}

To validate these theoretical insights obtained from a simple spring-mass model, we performed full-wave finite-element simulations with a realistic hinge design. We turn to a hinge geometry that favors bending over stretching \cite{bossart_oligomodal_2021, Note1} by combining a soft and a rigid material. Armed with these realistic hinges, our first step is to reproduce the spring-mass band structure in Fig.\ref{fig2}(c) using a finite-element eigenvalue study with Floquet-Bloch boundary conditions (conducted with COMSOL). The bi-material unit cell geometry is represented in the inset of Fig.\ref{fig2}(c) and \cite{Note1}. The theoretical and numerical low-energy phonon branches are in close agreement, with the exception of a mass gap appearing at the $K$ point. Such a gap is expected to arise as a result of the non-ideal hinges; its size is determined by the finite hinge stiffness. \rev{Above 1500 Hz, the predictions of FEM and theory start diverging; this reflects the fact that while both models share the same underlying mechanism, the internal structure of their rigid elements differs greatly. Once such elements start to deform, the spectra diverge. Here, our simple spring-mass model suffices to capture the dynamics below this threshold. This illustrates an important point, namely that several geometries can instantiate the same abstract mechanism, while their higher-frequency spectra generically differ. For further examples, see \cite{Note1}.}

Next, we explicitly demonstrate that the phase and group velocities around the anomalous cone are opposite, in a numerical experiment directly probing the refraction of a monochromatic gaussian beam at the interface between a standard isotropic medium and our nonlocally-resonant metamaterial. \rev{The metamaterial domain consists of an hexagonal tiling of $29\times 42=1218$ unit cells of the type depicted in the inset of Fig.2(c).} The momenta of the beams were selected to lie in the negative refraction region. The result, presented in Fig.\ref{fig2}(d), clearly evidences the negative refraction of the beam, consistent with the strong non-locality exhibited by the architected material \cite{forcella_causality_2017, belov_strong_2003}. \rev{We conducted similar negative refraction simulations \cite{Note1} for different frequencies, and found negative refraction behavior between $194$ and $317 Hz$, yielding a large relative bandwidth of $\frac{\Delta\omega}{\omega_{c}}=48\%$.} Therefore, anomalous cones can have important and drastic consequences on wave propagation, and be considered as a fundamental route to create metamaterials with broadband spatially dispersive effects.


To further solidify our discussion, we now seek experimental validation of these ideas. To do so, we 3D-printed a dual-material structure (Fig.\ref{fig3}(a) and \cite{Note1}) approximating the lattice of Fig.\ref{fig1}(bd). The white sections were printed with polyactic acid (PLA), whereas the orange sections forming the hinges were printed using a much more flexible thermopolyurethane (TPU) \cite{Note1}. We then excited elastic waves with pseudorandom noise in a range of $10-1250 Hz$ in our 61-cell hexagonal sample using a shaker and measured the internal rotation of each unit cell through laser vibrometry. The vibrometer was placed at a $26^{\circ}$ angle from the sample plane in order to favor in-plane velocity. Taking three measurement points per cell (green points in Fig.\ref{fig3}(e))  then allowed us to construct a measure of internal rotation, namely $\Omega:=2v_2-v_1-v_3$, where $v_1$ and $v_3$ were measured on the disconnected central triangles of the cell and $v_2$ in the center \cite{Note1}. Fig.\ref{fig3}(b-d) depicts spatial Fourier transforms of the results, superimposed with frequency contours obtained through the spring-mass model. Fig.\ref{fig3}(b) clearly indicates a concentration of energy at the $K$ and $K'$ points at $106 Hz$. As for Fig.\ref{fig3}(c), it shows how the anomalous cones open up with increasing frequency, progressively deforming into triangular contours. By $806 Hz$, the cones have merged and the resulting elliptic contours are still in agreement with the spring-mass model (Fig.\ref{fig3}(d)); they converge to the M point at $888 Hz$. This confirms the possibility of realizing nonlocally-resonant elastic metamaterials, with a large bandwidth: here, the anomalous branch extends from $106 Hz$ to $888 Hz$.


\begin{figure}[]
\includegraphics[width=\columnwidth]{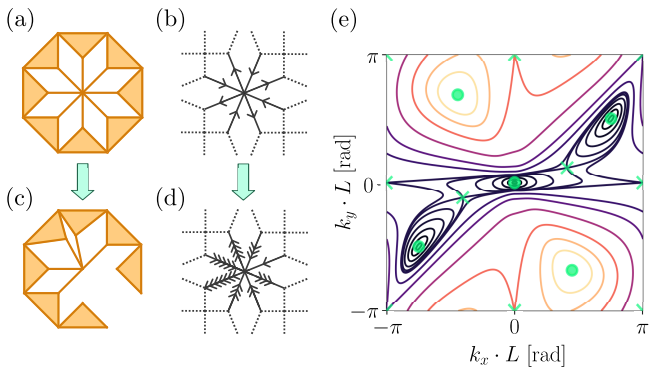}
\caption{\label{fig4} Inverse design of anomalous cones. (a) Octagonal unit cell with two anomalous cones. (b) Vertex model of associated mechanism. (c) Deformed unit cell, designed to exhibit an anomalous cone at an arbitrarily chosen location $(kx, ky)=(3\pi/4, \pi/2)$. (d) \rev{Graphical depiction of a vertex required to satisfy the target Floquet-Bloch boundary conditions}. (e) Isofrequency contours of the first dispersion branch for geometry (c). Extrema are indicated with green circles and saddle points with green crosses.}
\end{figure}

The examples we considered so far exhibited anomalous cones at highly symmetric points, and our method to find them, while practical and conductive to visual exploration, remained heuristic in nature. We want to go further and see whether inverse design is possible: given coordinates in \rev{k-space}, how do we design a geometry exhibiting an anomalous cone precisely at the specified location ?

\rev{Consider a combinatorially-designed} unit cell, \rev{for instance the one} of Fig.\ref{fig4}(a); it has an anomalous cone at the $M$ point \cite{Note1}. \rev{Through inverse design, we want to move this anomalous cone to an arbitrary location specified by the two components of the wavevector. Therefore, we start by allowing two geometric parameters in the unit cell to vary (the angles between the central links in Fig.\ref{fig4}(c)). This will generically frustrate the mechanism, by lifting symmetry-induced redundancies in the link constraints \cite{Note1}, and thereby} open up a small mass gap at the root of the anomalous cone. \rev{In order to compensate for this effect, we remove four constraints from the unit cell (here, three central bars and one triangle), as} shown in Fig.\ref{fig4}(c).

\rev{We then invert the problem of determining the location of the anomalous cone as follows: first, we enforce Floquet-Bloch boundary conditions by writing that the sum of arrows must vanish at every hinge connecting the unit-cell to its neighbours. These boundary condition equations can be collected in a matrix \cite{Note1}, whose kernel contains arrow configurations compatible with the target wavevector. Second, we compute the arrow rules for arbitrary values of the geometric parameters}, which yields a nonlinear system of equations relating the arrow weights to the unit-cell geometry \cite{Note1}. \rev{Finally, we combine the boundary conditions and the geometric equations to obtain the desired unit-cell parameters.}

\rev{More concretely, for $L\cdot(k_x, k_y)=(3\pi/4, \pi/2)$, the kernel of the arrow-conservation matrix contains a configuration approximated in Fig.\ref{fig4}(d) (the exact arrow weights are not integers). We can then insert these weights in our} system of \rev{nonlinear} equations for the geometric parameters \rev{\cite{Note1}}, \rev{which we invert} to obtain the desired metamaterial geometry\rev{, depicted in Fig.\ref{fig4}(c). With this, the inverse design procedure is complete and we can verify that the corresponding spring-mass model indeed exhibits an anomalous cone centered at $L\cdot(k_x, k_y)=(3\pi/4, \pi/2)$ in Fig.\ref{fig4}(e).} Since the system is symmetric under time-reversal, it naturally also presents another cone at $L\cdot(k_x, k_y)=-(3\pi/4, \pi/2)$.




The concept of nonlocally-resonant metamaterial, which unifies the present elastic metamaterials and interlaced wire media, also sheds new light on another type of metamaterial: roton metamaterials \cite{landau_helium, feynman_energy_1956,chen_roton-like_2021,iglesias_martinez_experimental_2021,kishine_chirality-induced_2020,chaplain_reconfigurable_2022,zhou_broadband_2022}. Although they are based on a different physical mechanism, namely long-range interactions, roton metamaterials also achieve wide domains of negative group velocity by leveraging a type of nonlocality. We have identified two ways in which our non-local resonances can open a mass gap and acquire a roton-like dispersion, namely through kinematic frustration or hinge design. Conversely, we surmise that electromagnetic rotons could be designed by perturbation of ideal interlaced wire media, either geometric or material in nature.

In conclusion, we introduced the notion of nonlocally-resonant metamaterials, wherein anomalous dispersion cones originate from arbitrary points in \rev{k-space}, associated to sample-spanning resonances. We have demonstrated that a directed-graph theory allows us to create such nonlocally-resonant metamaterials in elasticity, selecting the number and location of nontrivial zero-frequency modes in \rev{k-space}. We showed that the induced anomalous dispersion cones lead to bandgaps, slow sound and negative group velocities over a large bandwidth. We experimentally observed deep-subwavelength spatial dispersion in a 3D-printed nonlocally-resonant elastic metamaterial. Our findings provide a new toolbox for  broadband and low-frequency wave control leveraging spatial dispersion at the extreme.

\begin{acknowledgments}
A.B. and R.F. acknowledge funding from the Swiss National Science Foundation under Eccellenza Grant No. 181232 entitled "Ultracompact wave devices based on deep subwavelength spatially-dispersive effects".
\end{acknowledgments}








\nocite{*}

\bibliography{nonlocally_resonant}



\widetext
\clearpage
\maketitle
\begin{center}
\textbf{\large Supplemental Materials: Extreme Spatial Dispersion\\ in Nonlocally-Resonant Elastic Metamaterials}
\\
$\quad$\\
Aleksi Bossart and Romain Fleury

\textit{Laboratory of Wave Engineering, Ecole Polytechnique Fédérale de Lausanne, 1015 Lausanne, Switzerland}

(Dated: \today)
\end{center}
\setcounter{equation}{0}
\setcounter{figure}{0}
\setcounter{table}{0}
\setcounter{page}{1}
\makeatletter
\renewcommand{\theequation}{S\arabic{equation}}
\renewcommand{\thefigure}{S\arabic{figure}}
\renewcommand{\bibnumfmt}[1]{[S#1]}
\renewcommand{\citenumfont}[1]{S#1}

\color{black}
\section{Directed graph method}

\begin{figure}[!ht]
\includegraphics[width=\columnwidth]{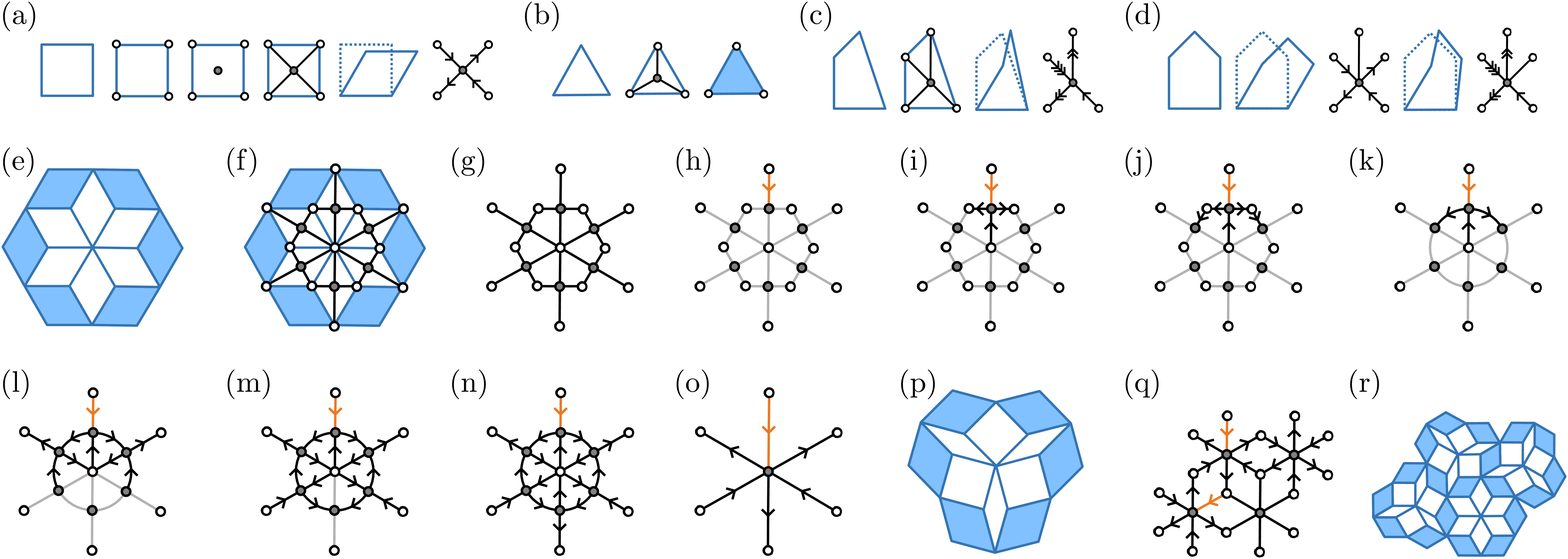}
\caption{\label{fS0} \color{black}Step-by-step description of the directed graph method. Blue lines correspond to rigid bars, filled polygons to rigid bodies and dotted blue lines to rigid bars in a reference configuration. Empty dots correspond to hinge vertices, full dots to linkage vertices, black lines to graph edges and arrows to angular deformations. Degrees of freedom chosen arbitrarily are depicted by orange links, whereas the black ones are fixed by kinematic rules as a consequence of this choice. The step by step diagrams also show grey links, whose motion remains to be determined during the process of filing up the graphs with arrows.\color{black}}
\end{figure}

Here, we supplement the description given in the main text by applying the directed graph method to several examples in a detailed manner. Let us start with a mechanism with one degree of freedom. It consists of four identical bars arranged in a square, depicted as black lines in Fig.\ref{fS0}(a). In order to draw the corresponding graph, we place four vertices on the hinges (white circles) and one vertex (black disk) inside the polygon formed by the bars, which we call a linkage. We then draw edges connecting the hinge vertices to the linkage vertex. Small deformations of this four-bar mechanism, such as the one depicted on the penultimate panel of Fig.\ref{fS0}(a), can then be represented on the graph by placing arrows on the edges. Increasing angles are drawn as arrows going from hinge vertices to linkage vertices, while decreasing angles go the opposite way. In the particularly simple situation of the square mechanism, the number of arrows per edge is fixed by symmetry: opposite corners see the same increase in angle, as represented in the last panel of Fig.\ref{fS0}(a). Furthermore, the total number of incoming arrows must equal the total number of outgoing arrows, since the sum of angles in a polygon is fixed.

In Fig.\ref{fS0}(b), we consider an even simpler case, namely a triangle of rigid bars, which yields a graph with three edges. What is the arrow-drawing rule in this case ? We know three relations among the inner angles of our triangle: they must sum to $\pi$ and satisfy the sine and cosine laws. Since this is a system of three variables and three equations, the inner angles are fixed and there is no mechanical degree of freedom. So, we do not draw any arrow. As a consequence, we can simplify graphs by erasing linkage vertices with three edges.

In Fig.\ref{fS0}(c), we consider a four-bar mechanism with unequal edges. As seen in the third panel of Fig.\ref{fS0}(c), angular deformations are no longer symmetric in this case. We therefore have to consider the trigonometric relations among the inner angles of our polygon, which consist of angle conservation and generalized sine and cosine laws. Since we focus on small deformations, we can linearize this system of equations and express three of the angles as multiples of the fourth. In this particular case, this yields the integer multiples represented by the integer number of arrows in the last panel of Fig.\ref{fS0}(c). Note, however, that in more complex geometries the arrow weights need not be integers, and one can simply indicate the weight of each arrow next to it.

We can also describe mechanisms with more than one degree of freedom. Take for instance the pentagon shown in Fig.\ref{fS0}(d): it has two degrees of freedom. The first, in which the upper corner of the pentagon is undeformed, is shown in the second panel of Fig.\ref{fS0}(d). It corresponds to the mechanism of a square. The second degree of freedom, shown in the last panels of Fig.\ref{fS0}(d), can be found by setting the angular deformation to zero on another corner. It corresponds to the mechanism of Fig.\ref{fS0}(c). Importantly, any linear combination of these arrow configurations is also admissible. Generally, for a n-sided polygon, one can still write three trigonometric relations, linearize them around a reference position and express $3$ angles in terms of the $n-3$ other. This translates into $n-3$ elementary arrow configurations that can then be combined at will.

We now consider the system of Fig.\ref{fS0}(e), which consists in an hexagonal arrangement of symmetric four-bar mechanisms. We carry out the graph-drawing procedure (Fig.\ref{fS0}(f)), yielding Fig.\ref{fS0}(g). Deforming the upper hinge of the unit cell then corresponds to drawing a single arrow in Fig.\ref{fS0}(h), which we highlight in orange to indicate that it was an arbitrary initial choice. The arrow-drawing rule of the linkage vertex then fixes three more edges (Fig.\ref{fS0}(i)). We then have our first encounter with arrow repartition at hinge vertices. At such vertices, the only constraint is that the sum of angles is conserved. To make this clear, consider a hinge vertex with $n$ edges; it corresponds to $n$ rigid bars meeting at a perfect hinge. These bars can rotate independently, and no trigonometric relationship constrains them besides angle conservation. In the case shown in Fig.\ref{fS0}(j), this constraint is sufficient to determine two more edges. In general, one can erase hinge vertices with two edges and directly connect the associated linkage vertices, as shown in Fig.\ref{fS0}(k). We can then apply the arrow rule of linkage vertices three more times, in Fig.\ref{fS0}(lmn), therefore exhausting all edges on the graph. We see that arrow conservation and kinematic rules are respected everywhere, in particular on the central hinge vertex, which shows that this mechanism is compatible. Furthermore, since we started by fixing a single angle, our unit cell has a single mechanism.

 Having worked through the grubby details, we can simplify matters greatly, by forgetting about the internal kinematics and drawing an effective linkage vertex with six edges and a single degree of freedom (Fig.\ref{fS0}(o)); the corresponding deformation mode of the system is depicted in Fig.\ref{fS0}(p). We can then tile the plane with this unit cell, building up the crystal considered in the main text (Fig.\ref{fS0}(q)) and see that we only need to fix two edges (highlighted) to determine the arrow configuration on the whole lattice. The metamaterial therefore has two global degrees of freedom, as discussed at length in the main text. One of these modes is shown in Fig.\ref{fS0}(qr), and is readily found to live at the K point. The strength of this graph method is its ability to combine different mechanical building blocks and then combinatorially determine the presence and symmetry of deformation modes in metamaterials composed of these building blocks.

\begin{figure}[!ht]
\includegraphics[width=\columnwidth]{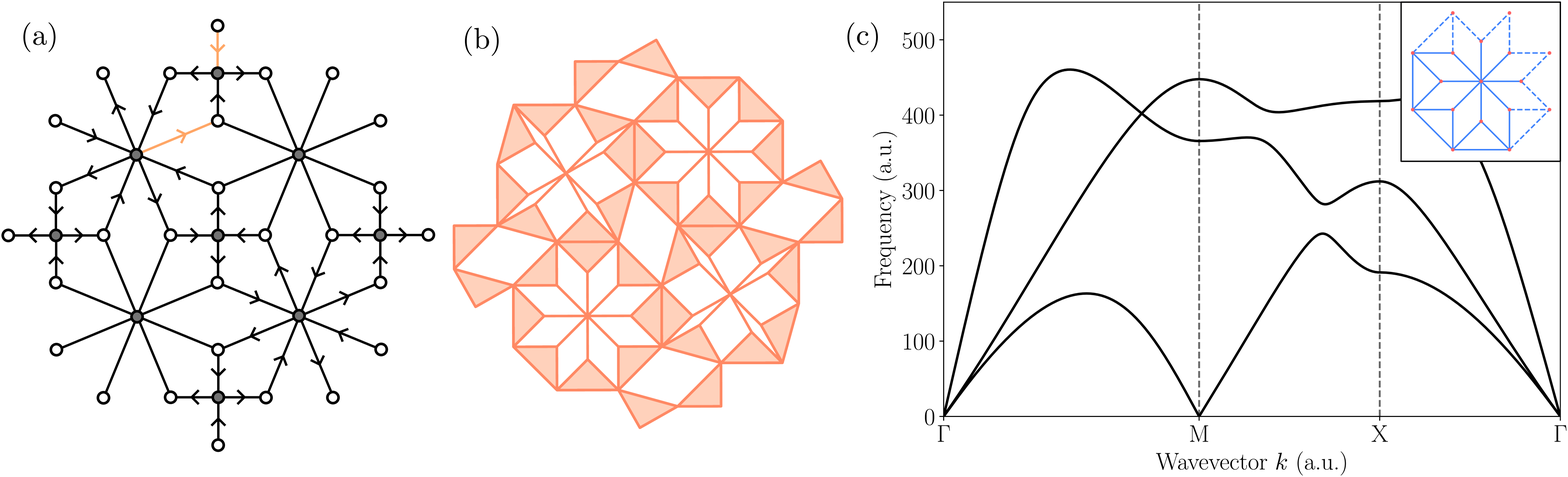}
\caption{\label{fS4} Alternative example of nonlocally-resonant metamaterial. \rev{(a) Directed graph with two global degrees of freedom; initial conditions are depicted as orange arrows. (b) Corresponding flexible metamaterial, with the mechanism of panel (a) actuated. }(c) Phonon spectrum in the symmetric rest position, depicted in the insert, with solid blue lines representing inner springs and dashed blue lines boundary springs. Point masses shown as red dots.}
\end{figure}

Let us now give another concrete example of combinatorial design, where we consider first a graph (Fig.\ref{fS4}(a)) built out of eight-legged and four-legged linkage vertices. We will deduce its static modes and the associated geometry that implements them. By fiddling around, we find that if the arrow rule is to alternate incoming and outgoing arrows of equal magnitude as we go around linkage vertices, the graph only has two global degrees of freedom. One of them is depicted in Fig.\ref{fS4}(a). Furthermore, these two degrees of freedom can be decomposed into one graph with $\Gamma$-point symmetry and another with $M$-point symmetry; this indicates that we would find anomalous cones at these locations, for structures with mechanisms realizing these arrow rules. We already have mechanisms with such arrow rules in our toolbox, namely the simple square linkage of Fig.\ref{fS0}(a) and the rosette-like mechanism of Fig.\ref{fS0}(o). Combining them yields the metamaterial of Fig.\ref{fS4}(b), where the mechanisms corresponding to the graph of Fig.\ref{fS4}(a) are actuated. In Fig.\ref{fS4}(c), we show the spectrum of a corresponding spring-mass model, which indeed exhibits extra anomalous cones at $\Gamma$ and $M$. In this example, standard waves hybridize with the anomalous ones and we have a complete band gap. 

\color{black}

\section{Impact of Pre-Twisting}

\begin{figure}[!ht]
\includegraphics[width=\columnwidth]{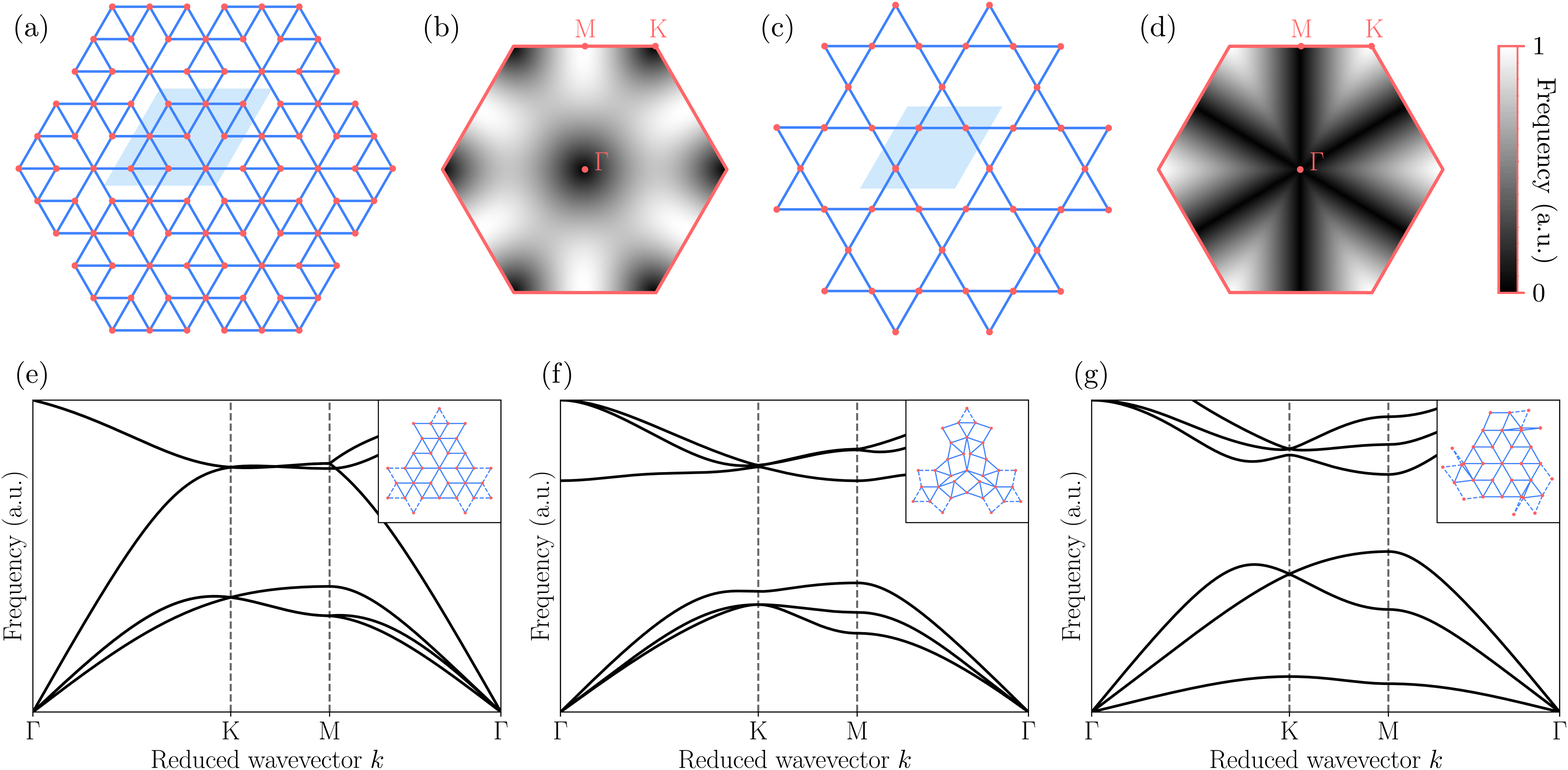}
\caption{\label{fS1} Spectra of \rev{several metamaterials. Solid blue lines represent springs, and point masses are shown as red dots. (a) Nonlocally-resonant metamaterial, with unit-cell highlighted in blue. (b) First band of the associated spectrum, with symmetry points indicated. (cd) Kagome metamaterial and its spectrum, exhibiting localized modes. (e-g) Spectra of pre-twisted versions of the nonlocally-resonant metamaterial of panel (a); the anomalous cone is folded to $\Gamma$ due to considering a larger unit cell.} The unit cells are represented in inserts.}
\end{figure}

\rev{We begin this section by contrasting the nonlocally-resonant metamaterial of Fig.\ref{fS1}(a) with the Kagome metamaterial, shown in Fig.\ref{fS1}(b). The first band of their phononic spectrum is represented in Fig.\ref{fS1}(c) and Fig.\ref{fS1}(a), respectively. As expected, the nonlocally-resonant metamaterial exhibits anomalous cones at the Brillouin zone corners; in contrast, the Kagome metamaterial presents lines of zero-frequency modes. Such lines are a signature of Guest-Hutchinson modes, which are also localized along lines in real space. As discussed in [20,21,36], pre-twisting the Kagome structure along one of its mechanical degrees of freedom removes these localized modes and leads to interesting topological effects. What would the effect of such pre-twisting on nonlocally-resonant metamaterials be? As a starting point, we note that} the geometry depicted in Fig.\ref{fS1}(a) actually hosts finite mechanisms. In other words, it is part of a two-parameter family of lattices that can continuously deform into each other; three examples are shown in the inserts of Fig.\ref{fS1}(e-g). In this section, we consider the band structure of such deformed lattices, which we term \textit{pre-twisted}. The first effect of pre-twisting the lattice is that it forces us to consider a larger primitive unit cell, in order to still be able to tile the plane periodically. In the limit of vanishing twisting, this induces the band-folding seen in Fig.\ref{fS1}(e). The main difference here is that all cones, anomalous and otherwise, are now located at the $\Gamma$ point. This may seem paradoxical; indeed, by a mathematical sleight of hand, all signs of negative group velocity have disappeared from the band structure. We can lift this paradox with the following thought experiment: consider an interface between our band-folded metamaterial and a standard isotropic medium, with positive group velocity. The normal component of the \rev{wavevector} is preserved as we cross the interface. To see what happens to the transverse component of \rev{wavevector}, we expand the periodic part of the Bloch wave-function, $\mathbf{u(r)}$, in the plane-wave basis of the isotropic medium, which shifts the traverse \rev{wavevector} by a multiple of the reciprocal \rev{basis} vector; in particular, we can obtain negative refraction if the expansion of $\mathbf{u(r)}$ is dominated by the appropriate plane-wave component. Therefore, the band folding does not remove the negative index property, it is just that the dominant energy of the eigenmode is now located in the second Brillouin zone.

Let us now consider Fig.\ref{fS1}(f); there, pre-twisting has opened a full band-gap by lifting the longitudinal waves to higher frequencies. Note that the low center frequency of said gap gives it a large relative bandwidth of $\frac{\Delta\omega}{\omega}=56\%$. For intermediate degrees of twisting, the longitudinal cone lifts continuously, not immediately inducing a complete band-gap; this only occurs at a critical twisting angle. Such gaps seem to occur generically for large amounts of pre-twisting, and can be continuously tuned through geometric parameters. We also remark that this type of pre-twisting demotes the crystallographic group of the metamaterial from $p6m$ to $p31m$ by removing hexagonal rotation symmetries while preserving mirror planes.

\begin{figure}[!ht]
\includegraphics[width=\columnwidth]{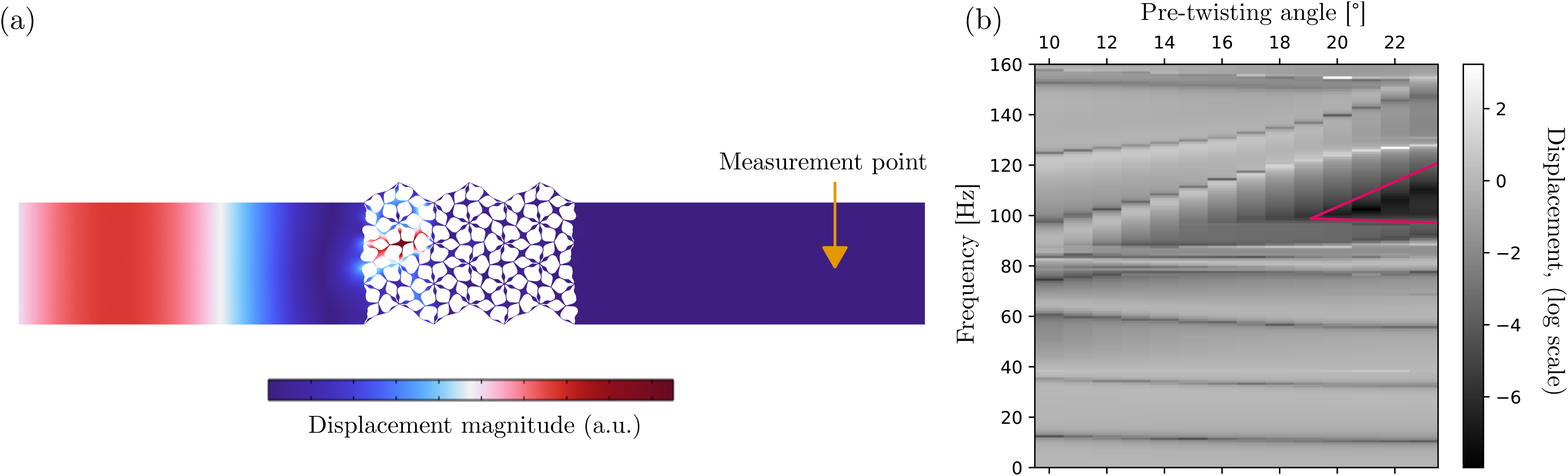}
\caption{\label{fS2} Impact of tunable band-gap on transmission through a metamaterial layer. (a) Geometry used in the simulation, with a $23^{\circ}$ pre-twist and a $108Hz$ longitudinal wave incoming from the left edge. (b) Resulting transmission of longitudinal waves as a function of the pre-twisting angle, with the theoretical bounds of the band-gap represented as solid red lines.}
\end{figure}

Fig.\ref{fS1}(g) demonstrates another interesting effect of pre-twisting, namely the impact it has on the speed of sound of anomalous waves. With very large twisting angles, the anomalous band becomes flat for all practical purposes. Admittedly, such extreme geometries are harder to realize in practice; nevertheless, this indicates a large range of tunability for the speed of sound. Here, the crystallographic group of the metamaterial goes from $p6m$ to $p6$, preserving rotational symmetries but not mirror symmetries. In Fig.\ref{fS2}(a), we show a finite-element simulation in which we placed three layers of pre-twisted metamaterial between two isotropic domains. The pre-twisting was chosen along the same mechanism as Fig.\ref{fS1}(g). We then conducted sweeps on frequency and twisting angle, with a plane wave coming from the left isotropic edge and a periodic boundary condition from top to bottom. Fig.\ref{fS2}(b) shows the resulting transmission to the right isotropic domain, with the theoretical bounds of the band-gap shown as red lines; we see that the critical angle along this mechanism is approximately $19^{\circ}$. This demonstrates that the effects of pre-twisting persist in full-wave simulations.

\color{black}
\section{Bandwidth of negative refraction.}

\begin{figure}[!ht]
\includegraphics[width=\columnwidth]{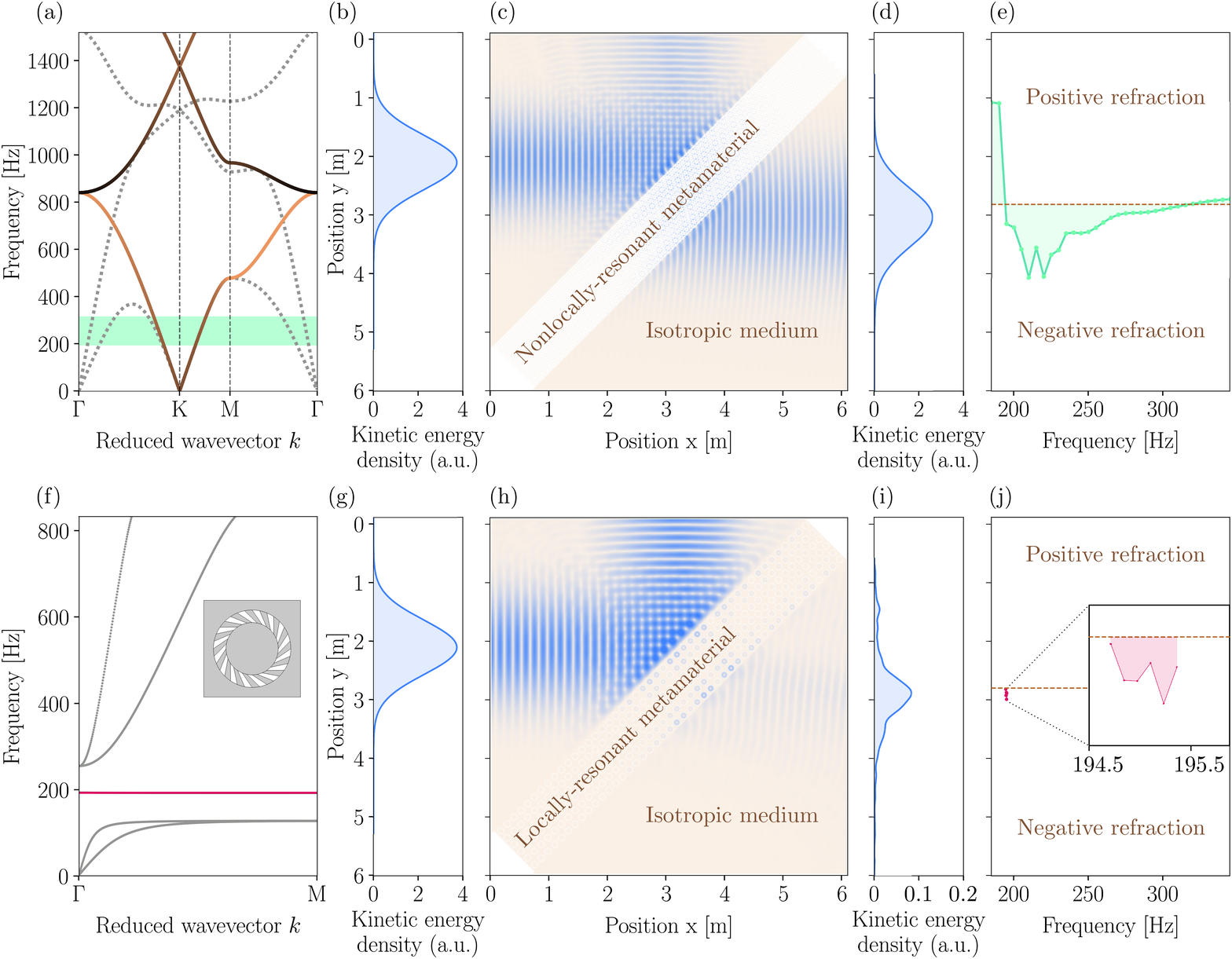}
\caption{\label{fSnegref} Bandwidth of negative refraction. (a) Band structure of the hexagonal unit-cell discussed in the main text (dashed lines) and pinned version (solid line). The frequency range of negative refraction is indicated in green. (b) Profile of a monochromatic gaussian beam. (c) Frequency-domain finite-element simulation with the aforementioned beam imposed on the left side, and perfectly-matched layers on all sides (not shown). (d) Profile of the beam on the right edge, after having been refracted twice. (e) Position of beam center for varying frequencies, with a dashed line indicating the transition from positive to negative refraction.\color{black} (f-j) Band structure and negative refraction for a traditional locally-resonant metamaterial. The unit cell is represented in the insert of panel (f). \color{black}}
\end{figure}

Here, we investigate the frequency-dependence of the negative refraction behavior discussed in the main text. In order to avoid secondary refracted beams, we pinned the unit cells of our metamaterial, thereby removing the cones corresponding to standard elastic waves. In particular, this will allow us to clearly isolate the negatively-refracted beam and fit it with a gaussian profile. The resulting band structure is shown in Fig.\ref{fSnegref}(a) (solid lines). As a reference, we reproduce the band structure of the unpinned structure (grey dotted lines). We then set up a frequency-domain FEM simulation, on the geometry of Fig.\ref{fSnegref}(c). A monochromatic gaussian beam (Fig.\ref{fSnegref}(b))is imposed on the left edge, goes through an isotropic medium, and is then refracted and reflected on a metamaterial interface. The boundaries of the isotropic medium are connected to perfectly-matched layers, to absorb outgoing waves. In Fig.\ref{fSnegref}(d), we plot the kinetic energy density of a beam exiting through the right boundary after having been refracted twice. We repeat the simulation for several frequencies and fit gaussian profiles on the aforementioned outgoing beams to estimate their center.  Comparing the results (Fig.\ref{fSnegref}(e)) with the brown dashed line that separates the domain of positive refraction, on top, from that of negative refraction, on the bottom, then allows us to ascertain the frequency range in which negative refraction takes place.  We see that our metamaterial is capable of negative refraction over a large relative bandwidth of $\frac{\Delta\omega}{\omega_c}=48\%$, from $194Hz$ to $317Hz$. Note that this negative refraction does not only depend on the properties of the metamaterial, but also on the elastic constants of the isotropic medium with which it interfaces. The isotropic medium we considered here has $E=7MPa$ and $\rho=600\frac{kg}{m^3}$; other choices can produce negative refraction bands all along the anomalous cone. \revtwo{For comparison (Fig.\ref{fSnegref}(f-j)), we carried out the same negative refraction experiment with a traditional locally-resonant metamaterial \cite{wang_two-dimensional_2016}; we used PLA as the matrix material, TPU as the hinge material and a $E=211GPa$, $\nu=0.49$ and $\rho=8950\frac{kg}{m^3}$ material for the central mass. This yielded a relative bandwidth of $\frac{\Delta\omega}{\omega_c}=0.33\%$.}

\color{black}

\section{Manufacturing}

\begin{figure}[!ht]
\includegraphics[width=\columnwidth]{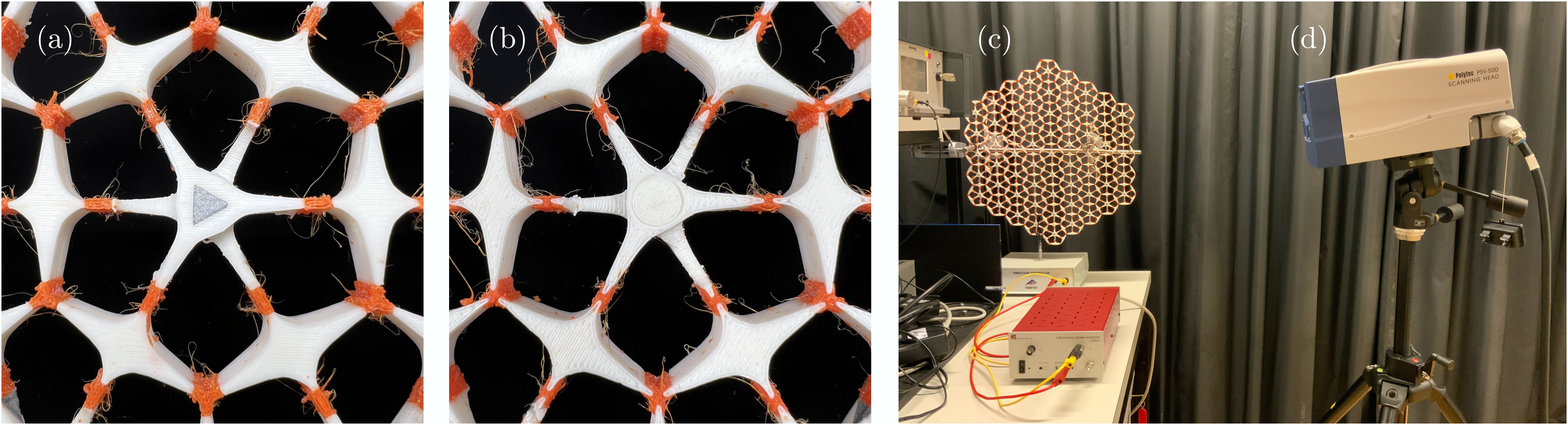}
\caption{\label{fS3} \rev{(ab) Closeups of the 3D-printed unit-cells within the sample. White sections were printed with PLA, whereas orange ones were printed in TPU. (c) Full} 3D-printed prototype. (\rev{e}) Laser vibrometer.}
\end{figure}

The unit cell of Fig.\ref{fS3}(a) was designed to assign a low energetic cost to deformations that correspond to the ideal mechanism discussed in the main text. To that end, the theoretical rigid springs were replaced by rigid volumes shaped to disfavor bending (\rev{white} regions in Fig.\ref{fS3}), whereas the hinges were implemented as soft and slender regions (\rev{orange in} Fig.\ref{fS3}). An air gap, shown in Fig.\ref{fS3}(b) and in the main text, allows the two central rigid structures to rotate relatively to each other. In order to favor bending over shearing and stretching, soft material was made to extend laterally on the rigid tips, as shown in Fig.\ref{fS3}(b). We 3D-printed the structure with a dual-nozzle Raise3D Pro2$^\text{®}$, using polyactic acid (PLA) as the rigid material, and Ninjaflex$^\text{®}$ thermopolyurethane (TPU) as the soft material. In order to increase contact area between the two materials, we alternated layers with deeply-penetrating TPU and standard ones, as shown in Fig.\ref{fS3}(b). The 61-cell structure (Fig.\ref{fS3}(d)) of the main text was printed in three parts, which were then connected along PLA interfaces using epoxy resin. Microbead-tape was added at measurement locations, as seen in Fig.\ref{fS3}(a), to improve retro-reflection. The vibrometer (Fig.\ref{fS3}(e)) was placed at a grazing angle of $26^{\circ}$. Vibrations were excited in the sample with a shaker, driven by an amplified sweep signal from the vibrometer ranging from $10Hz$ to $1250Hz$.

\section{Inverse design}

\begin{figure}[!ht]
\includegraphics[width=\columnwidth]{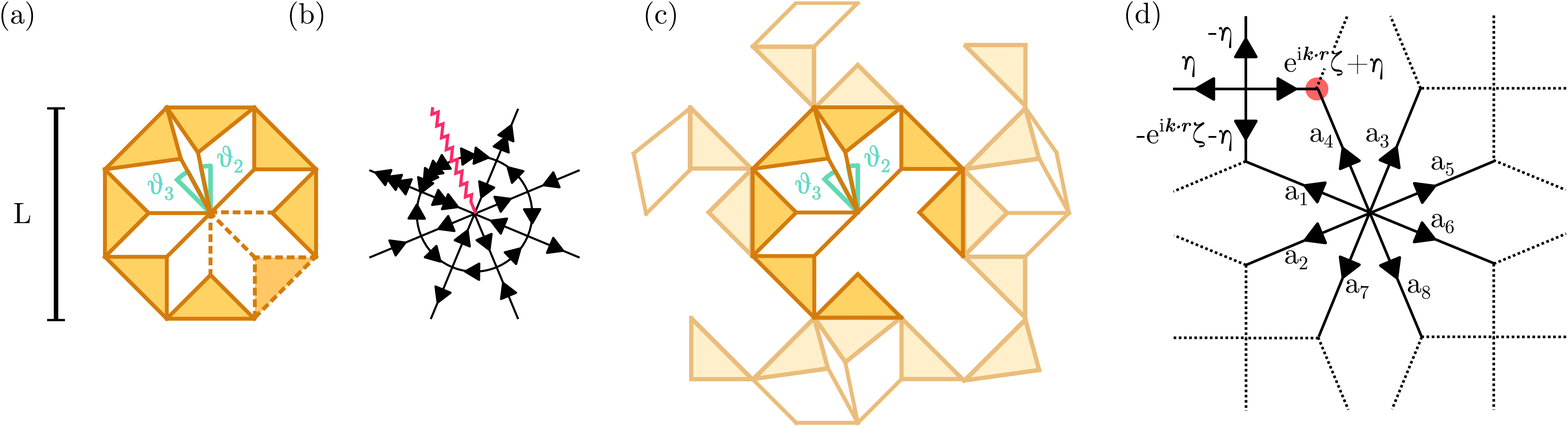}
\caption{\label{fS5} (a) Geometry of inversely-designed unit cell, with \rev{the tuning angles indicated in green. Connections depicted with dashed lines will later be removed. (b) Vertex model of the unit cell mechanism for generic tuning angles. The incompatibility of the constitutive linkages is indicated with a red line. (c) Unit cell in the metamaterial, with frustrating connections removed and} neighboring unit cells \rev{depicted} in reduced opacity. (b) Corresponding arrow weights on a directed graph, with $a5=-(\sqrt{2}-1)\zeta+\xi-a3$, $a6=-(\sqrt{2}+1)\xi-a4$, $a7=(\sqrt{2}-1)\zeta-\xi-a2$ and $a8=(\sqrt{2}+1)\xi-a1$.}
\end{figure}

The inverse design procedure described in the main text allows us to \rev{tune the unit cell geometry to induce} a prescribed shift of \rev{the} anomalous cone to an arbitrary position in \rev{k-space}. We now provide the computational details of \rev{this procedure}, applied to the case of the octagonal unit cell shown in Fig.\ref{fS4}(a). We know that this graph exhibits an anomalous cone at the M point, and our goal is to move it to a controlled location. \rev{To obtain the necessary degrees of freedom}, we allow \rev{two} geometric parameters to vary, \rev{namely} the two angles $\theta_2$ and $\theta_3$ indicated in Fig.\ref{fS5}(a). \rev{In Fig.\ref{fS5}(b), we see that trying to propagate arrows on the graph of this modified unit cell typically leads to incompatible configurations (red zigzag line). The mechanism becomes frustrated as we lift the octagonal symmetry, which was making some of the link constraints redundant. In order to compensate for this increase in the number of effective constraints, we remove four links from the unit cell (the dashed ones in Fig.\ref{fS5}(a)). We then consider the compatibility of the unit cell-mechanism with neighboring cells within the crystal, which are depicted in Fig.\ref{fS5}(c). They must deform together following the Floquet-Bloch boundary conditions, which we enforce by considering arrow conservation at hinges on the unit-cell boundary. For instance, consider the hinge highlighted in red in Fig.\ref{fS5}(d), where three edges meet. There, the sum of arrows is given by $a_4+e^{ik_y L}a_7+e^{ik_x L+ik_y L}\zeta+\eta=0$. We collect all such} arrow-conservation equations in the matrix
 
\begin{equation*}
    \left(
    \begin{array}{cccccccccc}
    0 & -1 & 0 & \cos (\tilde{k_x} ) & \cos (\tilde{k_x} ) & -\sin (\tilde{k_x} ) & -1 & 0 & 1 & 0 \\
    0 & 0 & 0 & \sin (\tilde{k_x} ) & \sin (\tilde{k_x} ) & \cos (\tilde{k_x} ) & 0 & -1 & 0 & 1 \\
    -1 & 0 & \cos (\tilde{k_x} ) & 0 & \cos (\tilde{\kappa} ) & -\sin (\tilde{\kappa} ) & \sqrt{2}+1 & 0 & -\sqrt{2}-1 & 0 \\
    0 & 0 & \sin (\tilde{k_x} ) & 0 & \sin (\tilde{\kappa} ) & \cos (\tilde{\kappa} ) & 0 & \sqrt{2}+1 & 0 & -\sqrt{2}-1 \\
    0 & \cos (\tilde{k_y} ) & 0 & -1 & -\cos (\tilde{\kappa} ) & \sin (\tilde{\kappa} ) & -\sqrt{2}-1 & 0 & 0 & 0 \\
    0 & \sin (\tilde{k_y} ) & 0 & 0 & -\sin (\tilde{\kappa} ) & -\cos (\tilde{\kappa} ) & 0 & -\sqrt{2}-1 & 0 & 0 \\
    \cos (\tilde{k_y} ) & 0 & -1 & 0 & -\cos (\tilde{k_y} ) & \sin (\tilde{k_y} ) & 1 &    0 & \cos (\tilde{k_x} )+\sqrt{2}-1 & \sin (\tilde{k_x} ) \\
    \sin (\tilde{k_y} ) & 0 & 0 & 0 & -\sin (\tilde{k_y} ) & -\cos (\tilde{k_y} ) & 0 & 1 & -\sin (\tilde{k_x} ) & \cos (\tilde{k_x} )+\sqrt{2}-1 \\
    \end{array}
    \right)
\end{equation*}

where the variables \rev{$\tilde{k_x}:=k_y L$, $\tilde{k_y}:=k_y L$ and $\tilde{\kappa}:=\tilde{k_x}+\tilde{k_y}$ correspond to the components of the wavevector at which we want to create an anomalous cone.} The components of the vectors upon which this matrix acts correspond to arrow weights in the vertex model. Our objective is to find the kernel of this matrix, which by design corresponds to \rev{abstract} sample-spanning mechanism\rev{s} with the spatial periodicities of the target Bloch wave. \rev{As an example, the arrow configuration depicted in Fig.3(d) of the main text was obtained in this way.} Such vectors have the form

\begin{equation}
    \left(
    \begin{array}{cccccccccc}
    a_1 & a_2 & a_3 & a_4 & \mathfrak{R}(\eta) & \mathfrak{I}(\eta) & \mathfrak{R}(\xi) & \mathfrak{I}(\xi) & \mathfrak{R}(\zeta) & \mathfrak{I}(\zeta)\\
    \end{array}
    \right),
\end{equation}

where the components are assigned as in Fig.\ref{fS5}(d). \rev{ Note that the $a_i$ parameters must have the same complex phase, since they move in concert; we can therefore set them all to be real and only consider the relative phases that may be picked up by the arrow weights $\eta$, $\xi$ and $\zeta$.} Having found two such vectors (our matrix has two more columns than rows), the second step of the method begins; we need to find geometric parameters that produce arrow rules compatible with the null vectors we identified. For the particular class of deformations we chose, only $a_1$, $a_2$, $a_3$ and $a_4$ depend on the geometric parameters. Since the norm of the arrow vector is arbitrary, we can normalize three of these arrow weights with respect to the first and obtain

\begin{equation*}
    \left(
    \begin{array}{c}
    a_2(w)/a_1(w)\\
    a_3(w)/a_1(w)\\
    a_4(w)/a_1(w)\\
    \end{array}
    \right)
    =
    \left(
    \begin{array}{c}
    \sin \left(\frac{\pi }{8}\right) (-\csc (\text{$\theta_2$})) \sec \left(\text{$\theta_3$}+\frac{\pi }{8}\right) \sin (\text{$\theta_2$}-\text{$\theta_3$})\\
    -\cos \left(\text{$\theta_2$}+\frac{\pi }{8}\right) \csc (\text{$\theta_2$}) \sin \left(\text{$\theta_3$}+\frac{\pi }{4}\right) \sec \left(\text{$\theta_3$}+\frac{\pi }{8}\right)\\
    \sqrt{\sqrt{2}+2} \cos \left(\text{$\theta_2$}+\frac{\pi }{8}\right) \csc (\text{$\theta_2$}) \cos \left(\frac{\pi}{8}-\text{$\theta_3$}\right) \sec \left(\text{$\theta_3$}+\frac{\pi }{8}\right)\\
    \end{array}
    \right),
\end{equation*}

in which the parameter $w$ encodes the linear combination of the two null vectors we found in the kernel of the boundary-condition matrix. This system can then be numerically solved for $w$, $\theta_2$ and $\theta_3$, yielding a geometry with the desired nonlocal resonance. This can be verified by removing the scaffolding and directly computing the band structure for this inversely-designed geometry, as we did in the main text.

\end{document}